# Self-starting Dynamics in All-fibre All-Normal-Dispersion Thulium Mamyshev oscillator

Dennis C. Kirsch[1,*], Kirill Grebnev[1], Alberto Rodriguez Cuevas[2], Mikhail E. Likhachev[3], Svetlana S. Aleshkina[3], M. V. Yashkov[4], Sonja Unger[1], Claudia Aichele[1], Adrian Lorenz[1], Auro Perego[2] and Maria Chernysheva[1]

[1]*Leibniz Institute of Photonics Technology, Jena, Germany*
[2]*Aston Institute of Photonics Technologies, Aston University, Birmingham, UK*
[3]*Prokhorov General Physics Institute of the Russian Academy of Sciences, Dianov Fiber Optics Research Center*, Moscow, Russia
[4]*Institute of Chemistry of High Purity Substances of the Russian Academy of Sciences, Nizhny Novgorod, Russia*

[*]*dennis.kirsch@leibniz-ipht.de*

**Abstract**

Ultrashort pulse formation from noise represents a fundamental self-organisation process in nonlinear dissipative systems and remains central to ultrafast photonics. Mamyshev oscillators offer a particularly valuable platform for investigating this phenomenon because they do not rely on conventional saturable absorbers. Instead, pulse formation is governed by self-phase modulation in normal-dispersion fibres combined with periodic offset spectral filtering. Achieving self-starting pulse formation in these systems is challenging, particularly at longer wavelengths, due to limited normal-dispersion components and the complex gain dynamics. This work reports, to the best of current knowledge, the first self-starting, all-fibre, all-normal-dispersion Thulium-doped Mamyshev oscillator operating near 1.9 µm. The cavity employs a compact Fabry–Perot design incorporating a dispersion-engineered Thulium-doped gain fibre, a highly nonlinear passive normal-dispersion fibre, and a pair of chirp-free broadband fibre Bragg gratings. The oscillator self-starts without external seeding or active modulation and stabilises noise-like pulse generation regime at a fundamental cavity repetition rate . Real-time measurements and numerical simulations reveal the build-up pathway from noise through transient multi-pulsing and pulse competition to a stationary noise-like envelope. Our results show that both tailored laser components and gain-medium-specific dynamics are essential for enabling self-starting ultrashort pulse generation in Mamyshev oscillators and for extending these laser concepts beyond the near-infrared, thereby facilitating the development of compact and robust shortwave infrared (SWIR) sources and new regimes of ultrafast self-organisation.

## Introduction

The formation of coherent structures from noise is a universal feature of nonlinear dissipative systems across physics, chemistry, and biology. Ultrafast fibre lasers provide a particularly powerful platform for studying this self-organisation, while simultaneously enabling a wide range of practical photonic technologies. Although ultrashort pulse generation through locking the phases of several cavity modes (mode-locking) has been investigated for decades (ref. 1,2), the pathways by which complex nonlinear laser cavities evolve from noise to stable pulsed states remain an active area of fundamental and applied interest.

Within this context, Mamyshev oscillators have become a particularly appealing platform for exploring pulse self-organisation dynamics in ultrafast fibre lasers. Their operating principle relies on the interaction between self-phase modulation (SPM)-induced spectral broadening in the normal dispersion regime and offset spectral filtering, which creates a steep, intensity-dependent transmission profile and enabling pulse formation without a traditional saturable absorber (ref. 3,4). In normal-dispersion cavities, phase locking of the modes has been understood as resulting from dissipative Faraday instability, where spectral reshaping and periodic, alternating frequency-detuned spectral filtering produce parametric gain bands that initiate the temporal localisation of pulses (ref. 5,6). Crucially, in contrast to many other mode-locking schemes, this mechanism in net-normal-dispersion cavities can tolerate very large accumulated nonlinear phase shifts (over $60\pi$), making Mamyshev oscillators highly promising for high-energy ultrafast operation. Consequently, this architecture has become a vital platform for achieving megawatt peak powers (ref. 7,8), high harmonic generation with repetition rates reaching up to a few gigahertz (ref. 9), and pulse durations of less than 60 femtoseconds.

Despite rapid advances in designing Mamyshev oscillators, successful ultrafast generation has primarily focused on the 1.0- and 1.55-µm bands, where Yb-, Er-, and Raman-based fibre laser systems benefit from readily available normal-dispersion fibre gain media, passive fibres, and fibre-compatible laser components (ref. 5,7,10). Extending this approach into the short-wave infrared (SWIR) range, from 1.7 to 2.5 µm, is highly promising for applications, spanning the precision materials processing, environmental sensing, ranging, and biophotonics (ref. 11-13). However, operation within this wavelength region is inherently more challenging because standard silica fibres exhibit significantly stronger anomalous group-velocity dispersion and decreased effective Kerr nonlinearity. Specifically, at 2 µm, the anomalous dispersion is approximately four times higher than in the near-infrared, while the increased mode area reduces the effective nonlinear phase accumulation by a similar factor. This makes it more difficult to sustain monotonic SPM-driven spectral broadening and increases the probability of pulse breathing and breakup. As a result, the design space for self-starting, all-fibre Mamyshev oscillators in the SWIR remains severely constrained.

In Thulium (Tm)-doped fibre Mamyshev oscillators, these constraints have so far led to unfavourable compromises in cavity design and operation. Early implementations typically relied on free-space spectral filters and external bulk compressors (ref. 14, 15), while still requiring either external seeding (ref. 15, 16) or dynamic adjustment of the filter separation (ref. 17) to achieve stable operation. In non-polarisation-maintaining designs, the inclusion of polarisation-sensitive elements such as waveplates and beam splitters further introduced the possibility of nonlinear polarisation evolution or intensity-dependent loss, making it difficult to attribute pulse formation solely to the Mamyshev mechanism. Recently, all-fibre-integrated Mamyshev oscillators have achieved sub-400-fs pulse durations after external compression (ref. 16,17). Nonetheless, the cavities still contained anomalous-dispersion fibre sections, especially in the gain medium or fibre-based components, which complicated pulse evolution and reduced the clarity of the underlying mode-locking mechanism.

Recent progress in dispersion-engineered speciality fibres and fibre-compatible components has begun to alter this situation (ref. 18, 19). In particular, Tm-doped germanate-rich core fibres can provide intrinsically normal dispersion along with efficient gain around 2 $\mu$m (ref. 20,21). Complementary, waveguide dispersion can be engineered, for example, by introducing a trench in a fibre cladding via Fluorine doping, enabling near-zero or highly normal dispersion with enhanced nonlinearity in the same spectral region (ref. 22). The most advanced fabrication approaches involve creating all-solid microstructured and hybrid multi-glass fibres, which enable flat, broadband normal dispersion across the 1.55–2.5 µm range (ref. 23). These advances make it possible to design cavities where both amplification and spectral broadening occur predominantly in the normal-dispersion regime, thereby avoiding the anomalous-dispersion sections that have complicated previous Tm-doped Mamyshev systems.

In this work, we demonstrate a self-starting all-fibre, all-normal-dispersion Tm-doped Mamyshev oscillator operating at 1.9 µm. The cavity is enabled by dispersion-engineered active and passive optical fibres, together with broadband flat-top fibre Bragg gratings (FBGs) that provide chirp-free offset spectral filtering. By combining real-time measurements with numerical modelling, we reveal the build-up pathway from noise to stable operation in a noise-like pulse generation regime. The oscillator delivers a maximum average output power of 50 mW, corresponding to 27 nJ pulse energy confined within a ~1.77-ns envelope at the fundamental repetition rate. These results establish all-normal-dispersion all-fibre Tm-doped Mamyshev oscillators as a promising platform for studying ultrafast self-organisation and for developing compact low-coherence SWIR sources.

## Results

### All-normal-dispersion Tm-doped fibre Mamyshev cavity design

Figure 1a illustrates the architecture of the self-starting all-fibre Tm-doped Mamyshev oscillator. Unlike previously reported 2-µm Mamyshev systems, the current cavity is designed to operate entirely within the normal-dispersion regime by omitting unnecessary fibre-based laser components, such as wavelength-division multiplexers, isolators, or couplers. These components, in addition to causing undesirable insertion losses, would reintroduce anomalous dispersion and complicate the mode-locking process. Therefore, the oscillator is built in a simple linear configuration using only three functional elements: a dispersion-engineered Tm-doped gain fibre, two segments of highly nonlinear, near-zero-normal-dispersion passive fibre, and an offset pair of broadband, flat-top FBGs.

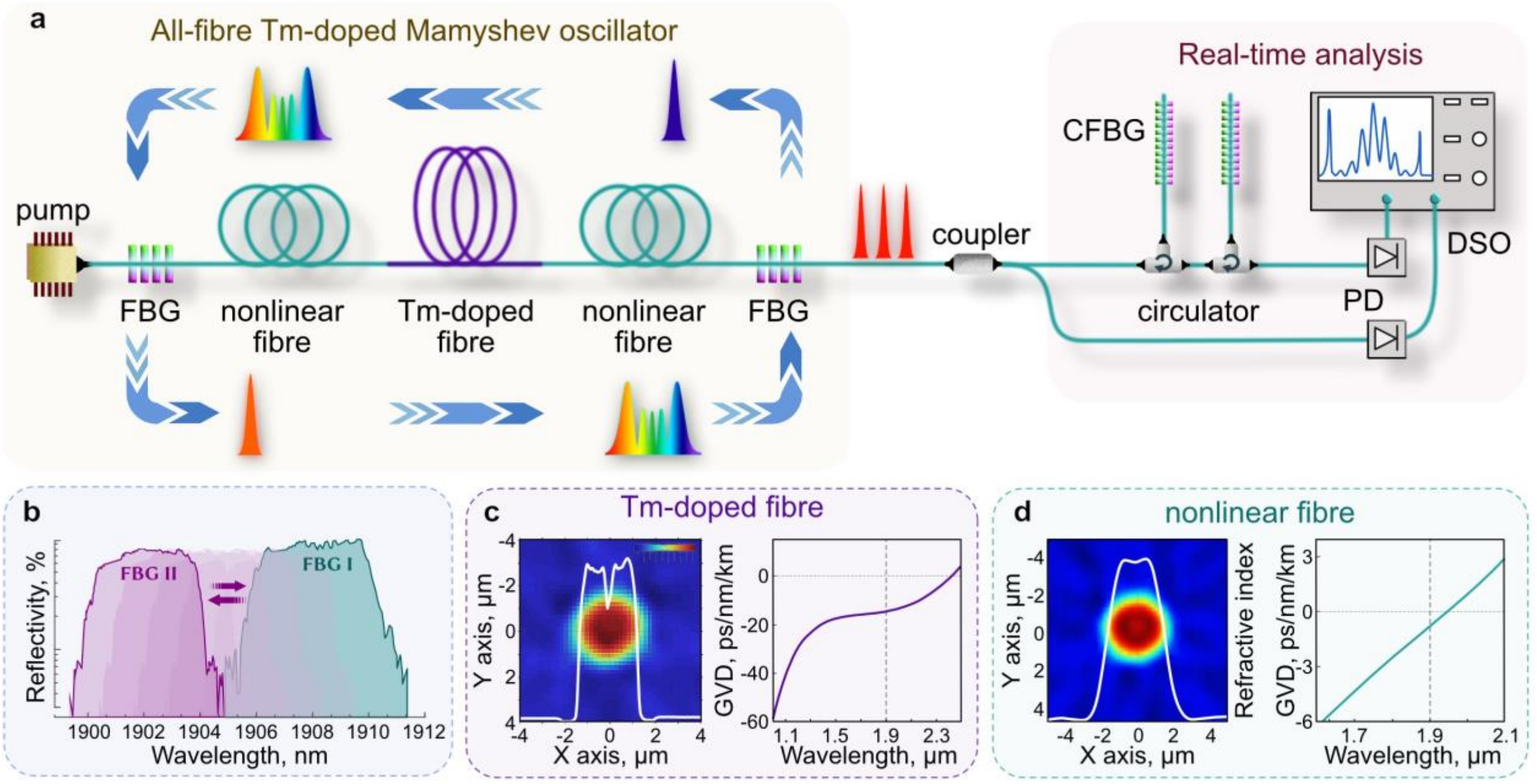

**Figure 1: Self-starting all-normal dispersion all-fibre Tm-doped Mamyshev oscillator. (a)** Schematic of laser cavity; **(b)** reflectance profile of filters and their tuneability; refractive index profile and dispersion of gain **(c)** and passive **(d)** nonlinear fibres.

Both passive and active speciality fibres feature small cores and high $GeO_2$ concentrations, which enhance the refractive-index contrast and shift the zero-dispersion wavelength beyond 2 $\mu$m (Fig. 1 c,d). Consequently, both fibres exhibit weakly normal, relatively flat dispersion over the operating range from about 1.6 to 1.9 µm. In this configuration, a nearly linear chirp is built up due to the combined effects of normal group-velocity dispersion and SPM. The offset FBGs at each round trip reinforce the Mamyshev regenerator transfer function, reset the temporal waveforms, and establish a dissipative balance that prevents wave-breaking and allows energy scaling (ref. 24,25).

A notable feature of the cavity is the absence of polarisation-sensitive elements or controllers. This excludes the contribution of nonlinear polarisation evolution to the mode-locking mechanism and allows the observed self-starting behaviour to be attributed solely to the interplay of gain, dispersion, nonlinearity, and offset spectral filtering. The total cavity length is 56 m, corresponding to a repetition rate of 1.8 MHz, while the net cavity group-delay dispersion is 0.44 $ps^2$ (excluding the dispersion of FBGs). Additional details on the laser cavity and component design are provided in Materials and Methods.

### Self-starting and steady-state noise-like pulse generation

The oscillator self-starts at a pump power of ~1 W when the spectral separation between the central wavelengths of FBGs is set between 5.5 and 6.5 nm. While near the threshold, a single weak mechanical perturbation is required to trigger pulse build-up, fully seed-free self-starting can be achieved at higher pump powers around 3 W (with a moderate FBG offset of 5.6 nm) without external modulation. Throughout the explored operating range, with pump powers from 1 to 4.1 W and an FBG's central wavelength offset of 5 to 8 nm, the laser consistently operates at the fundamental cavity repetition rate and does not enter the harmonic mode-locking regime. Once stable pulse generation is attained, the system remains robust against environmental disturbances, despite comprising only non-polarisation-maintaining fibres.

Characteristic steady-state pulse trains in the time domain are illustrated in Fig. 2(a,b). The output exhibits a stable train of rectangular pulse envelopes with only minor aperiodic peak amplitude fluctuations.

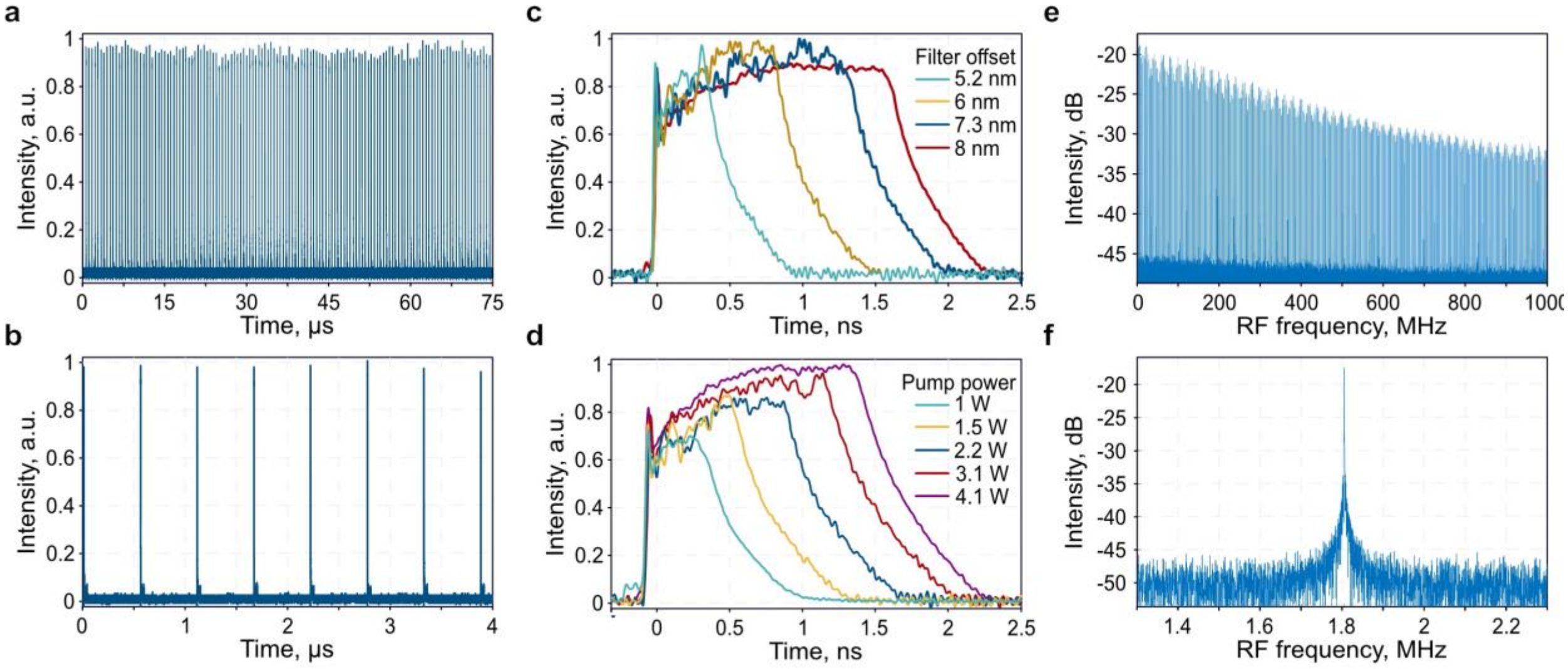

**Figure 2: Temporal characteristics of the generated pulse train in the all-fibre Tm-doped Mamyshev oscillator. (a,b)** Pulse train and zoomed-in pulses; variation of pulse duration with the alteration of filters mismatch at 2.5 W pump power **(c)** and pump power at fixed filter offset of ∼6nm **(d)**; **(e,f)** radio-frequency spectrum and zoomed-in fundamental frequency peak at pump power 1 W and filter separation ∼6nm.

Over a ∼3-ms acquisition window, the pulse-to-pulse intensity root-mean-square (RMS) remains at 1.06%, demonstrating stable operation once the cavity reaches its stationary state. The pulse envelope duration can be adjusted by tuning both the filter separation and the pump power, as shown in Fig. 2(c,d). The increase in the FBG offset from 5 to 8 nm broadens the envelope from 470 ps to 1.77 ns at a fixed pump power of 2.5 W. Similarly, the pump power from 1 to 4.1 W at a fixed ∼6 nm filter mismatch produces a comparable broadening trend. Overall, at maximum pump power, the oscillator delivers a 50-mW average output power, corresponding to 27 nJ pulse energy. The RF spectrum in Fig. 2(e,f) confirms stable operation at the 1.8 MHz fundamental repetition rate, with weak sidebands in the GHz-offset region attributable to envelope-duration jitter rather than harmonic pulsation. It is important to note that the RF spectrum is obtained via a fast Fourier transform of the oscilloscope trace. Therefore, its signal-to-noise ratio is limited by the instrument's dynamic range (effective number of bits 5.5).

The average output optical spectrum is shown in Fig. 3(a). As the pump power increases, the spectral bandwidth broadens from 12 to 17 nm while maintaining an approximately triangular profile. At the highest pump powers, a weak Raman shoulder appears near 2.0 μm. The spectral notch around 1.9 μm originates from the out-coupling FBG, as illustrated in the inset in Fig. 3(a), and is thus an inherent cavity feature rather than a sign of internal spectral instability.

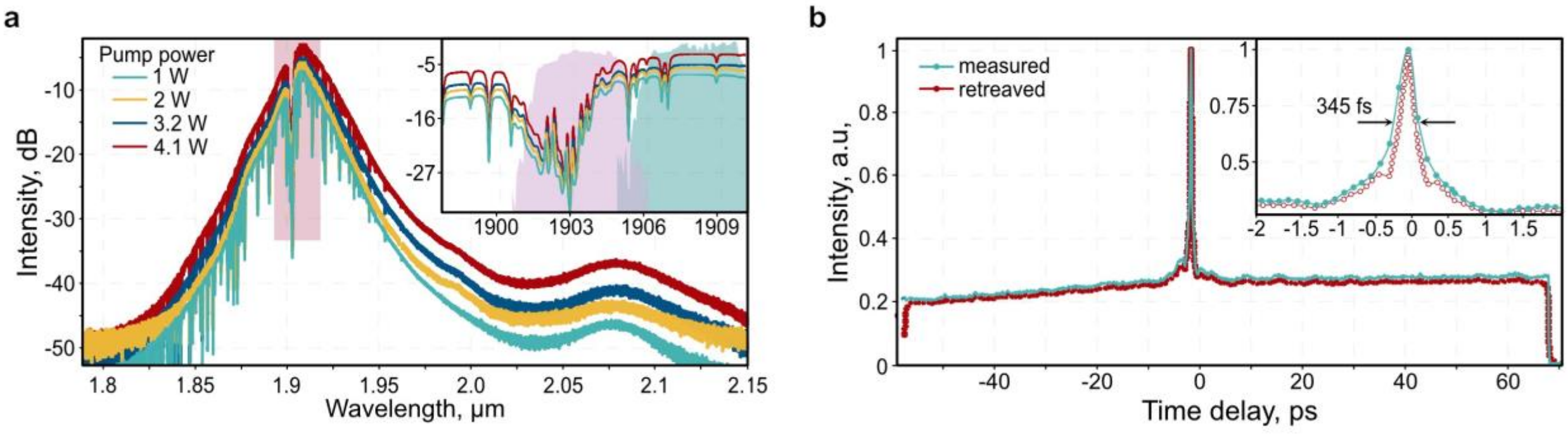

**Figure 3: Averaged output characteristics of noise-like pulse generation: (a)** Optical output spectrum at different pump powers. Inset: zoomed-in central part featuring schematic reflection bands of FBG filters; **(b)** Characteristic autocorrelation trace, recorded at 6 nm filters mismatch and 1.5-W pump power, demonstrating broadband intensive pedestal over 120 ps. Inset: Zoomed-in 345-fs coherence peak.

The autocorrelation trace in Fig. 3(b) shows a broad pedestal extending across the entire measurement window, with a narrow coherence spike at zero delay. This is characteristic of noise-like pulses. The coherence spike corresponds to a sub-pulse coherence time of approximately 0.34 ps, while the broad pedestal represents a nanosecond-scale envelope filled with rapidly fluctuating pulse substructures. Since the pulse positions fluctuate randomly within the bunch faster than the photodiode response time, the oscilloscope records them as a solid rectangular envelope, consistent with Fig. 2(c,d). The relatively strong contrast between the pedestal and the coherence spike in the autocorrelation trace suggests a low degree of coherence among the sub-pulses, with stochastic variations and densely packed sub-pulses. Alongside the rectangular temporal envelopes observed directly on the oscilloscope, these data confirm that the stationary regime is noise-like pulse generation. This conclusion is further supported by numerical simulations, which also do not tend towards a conventional coherent single-pulse generation regime.

To clarify the physical origin of this operating regime, we modeled the cavity using a generalised nonlinear Schrödinger equation framework that iteratively describes the evolution of the slowly varying field envelope over successive round trips. In each round trip, the field propagates through the passive fibre, the Tm-doped gain fibre, and the second passive fibre section, after which the cavity boundary conditions are imposed by multiplying the field spectrum with the reflection profile of one of the used FBGs. Furthermore, the field is back-propagated, and its spectrum is multiplied by the reflection profile of the other FBG. The model incorporates dispersive and nonlinear propagation in both passive and active fibres, gain saturation in the Tm-doped segment, cavity loss, and spectral filtering imposed by the two offset FBGs. Although simplified, this approach has previously been shown to capture the dynamics of linear-cavity Mamyshev oscillators (ref. 5) and, in the present case, reproduces the key experimentally observed features of both the stationary and transient regimes in semi-quantitative agreement.

The simulated temporal and spectral evolution is shown in Fig. 4. Starting from low-intensity nanosecond-scale fluctuations, the field develops into a stable broad envelope with a correspondingly wide spectrum. Importantly, the model does not converge towards a conventional coherent single-pulse solution. As it propagates through near-zero or weakly normal-dispersion highly nonlinear optical fibres, the pulse envelope accumulates strong chirp and spectral broadening due to intense SPM, while periodic filtering at FBGs truncates the broadened spectrum on each round trip. Under these conditions, the cavity supports a stable broad envelope with random internal structure but does not converge into a fully coherent pulse. The steady-state pulse characteristics shown in the side panels of Fig. 4(a,b) closely resemble the experimentally recorded nanosecond-scale noise-like pulse envelope, along with a broad optical spectrum containing the characteristic spectral dip imposed by the cavity filters.

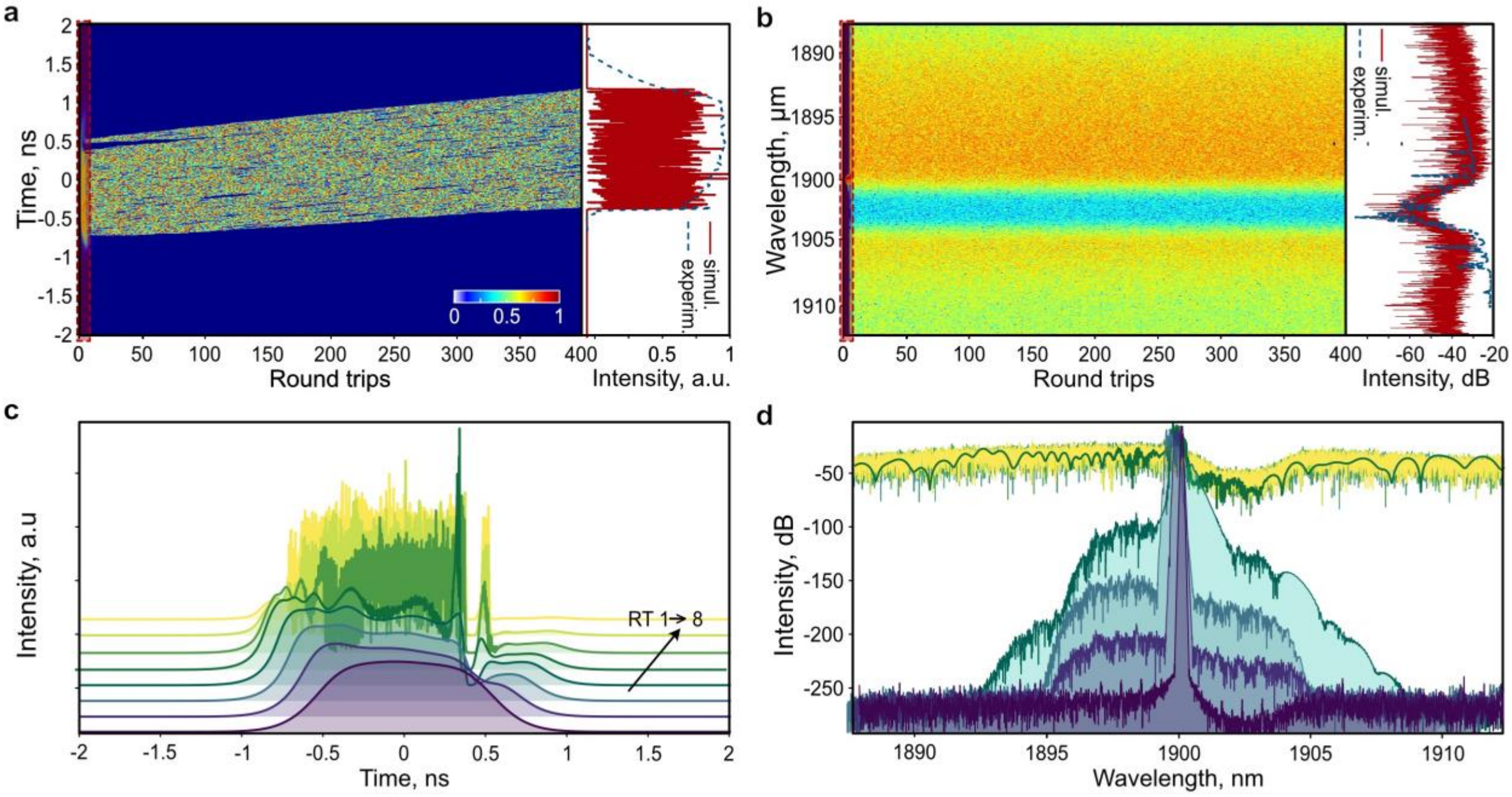

Figure 4: **Simulation results:** Build-up evolution of a) noise-like pulse and b) power spectrum over 400 cavity round-trips. Side panels: agreement between numerical simulations and experimental measurements of the steady-state noise-like pulse and its spectrum, showing the grating dip. Detailed formation dynamics over the first 10 round-trip in c) temporal and d) spectral domain.

**Build-up dynamics in the all-normal-dispersion Tm-doped fibre Mamyshev oscillator**

While steady-state measurements and simulations have established the nature of the final noise-like pulse operating regime with excellent agreement, the averaged experimental measurements do not reveal the pathway by which the cavity reaches this regime. To confirm the simulated dynamics experimentally, we performed round-trip-resolved measurements of the regime initialisation from noise in both the temporal and spectral domains using real-time dispersive Fourier transform (DFT) measurements (ref. 26).

To date, the DFT technique has enabled direct observation of a wide variety of complex nonlinear dynamics and ultrafast phenomena (ref. 26-28), including pulse build-up and transient dynamics. However, real-time studies of Mamyshev oscillators often failed to reveal the intrinsic self-organisation and nonlinear dynamics that are solely inherent to spectral reshaping in normal-dispersion media and to periodic filtering. The earlier demonstrations were largely obscured by the parameters of externally injected seed pulses, used to initiate ultrashort pulse generation (ref. 29-31).

Figures 5 and Supplementary Movie S1 show the full build-up process recorded at a pump power of 1.58 W and an FBG separation of about 5.6 nm. During the first roughly 300 round trips, the round-trip-resolved intensity evolution in Fig. 5(a) exhibits irregular amplification and decay of the background, along with repeated nucleation of low-intensity multi-pulse structures. Around RT 290, a transient multi-pulse pattern becomes apparent, and by RT 297, a higher-intensity two-pulse state has formed. Over the subsequent ~300 round trips, these pulses compete, with the weaker one gradually being suppressed until only a single dominant envelope remains. By RT 635, the cavity attains a stable single-envelope state that persists for the rest of the acquisition period.

The single-shot spectral evolution shown in Fig. 5(b) reflects this temporal process. In the initial phase, the spectrum fluctuates near the noise floor without forming a stable filtered profile. During the transient two-pulse stage, the main spectral component shifts slightly towards longer wavelengths and gradually overlaps more with the FBG reflection band. Once pulse competition ends, the remaining envelope broadens into the filter-limited stationary spectrum typical of steady state. Importantly, the DFT traces do not display stable high-contrast fringes linked to a fixed sub-pulse spacing, which supports the interpretation that the final state is noise-like rather than a phase-locked pulse molecule or a coherent multi-pulse state.

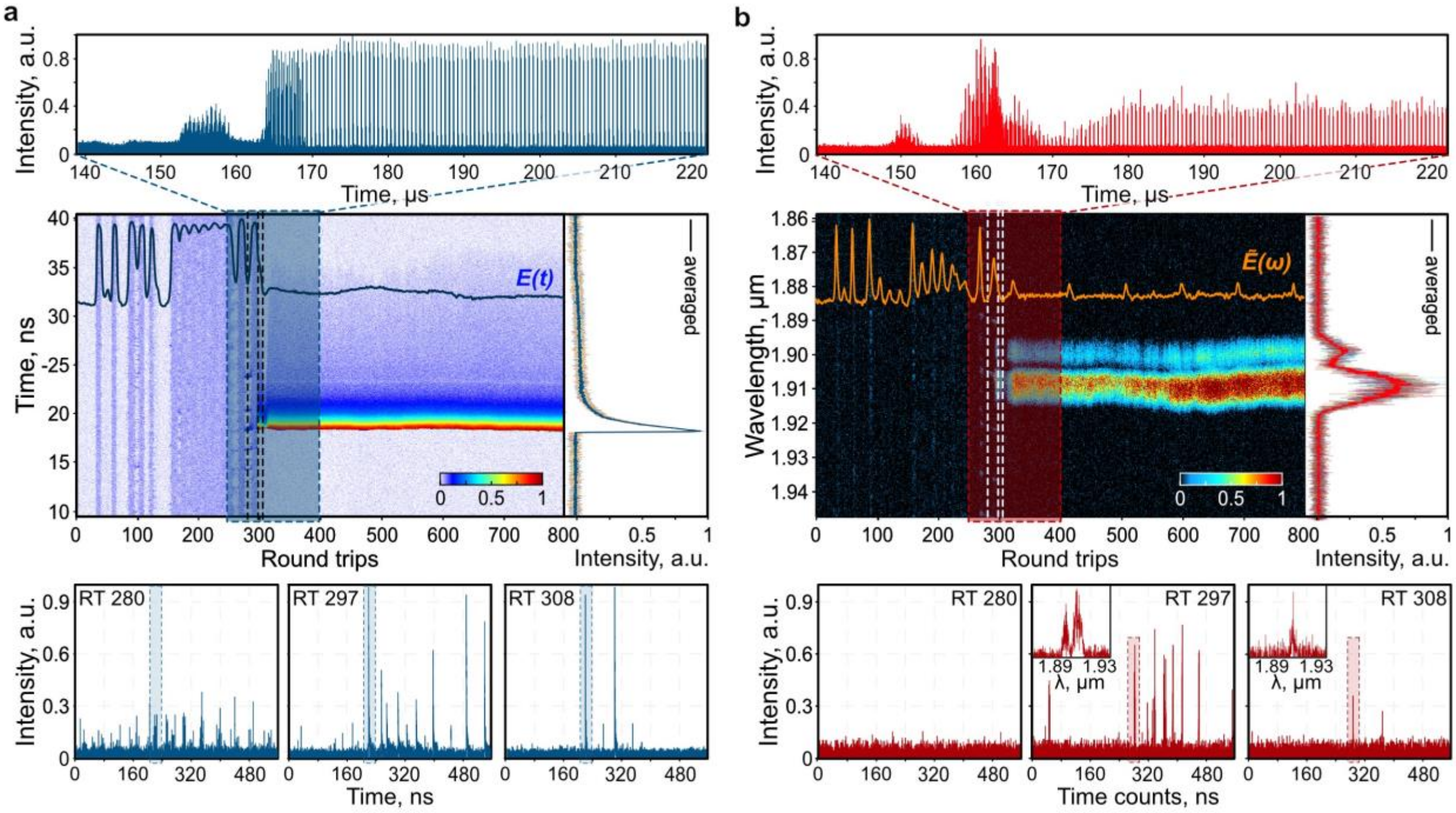

**Figure 5: Pulse build-up process in the Tm-doped fibre Mamyshev oscillator: (a)** Round-trip resolved intensity evolution in the time domain. Side panel: Steady-state pulse profile averaged over 10 round-trips; **(b)** Single-shot spectral evolution. Side panel: Averaged over 10 round-trip DFT spectra in the steady-state.

The recorded build-up dynamics differ from the scenarios of dissipative-soliton start-up (which typically proceeds via a Q-switched burst followed by pulse break-up) (ref. 32-35) and from pulse dynamics in normal-dispersion laser cavities, as demonstrated in (ref. 36,37), which feature oscillatory structures during the mode-locking transition. Neither does it resemble the externally seeded start-up dynamics commonly studied in Mamyshev oscillators. Instead, the present cavity evolves through stochastic background amplification, transient multi-pulsing, and pulse competition before settling into a single noise-like envelope at the fundamental repetition rate.

**Discussion**

We have demonstrated a self-starting, all-fibre, all-normal-dispersion Tm-doped Mamyshev oscillator operating around the 1.9 µm band. The cavity integrates a dispersion-engineered Tm-doped gain fibre, highly nonlinear near-zero-normal-dispersion passive fibre, and broadband flat-top FBGs within a minimal Fabry–Perot architecture. This design eliminates several limitations seen in earlier 2-µm Mamyshev oscillators, such as the need for external seeding, active filter modulation, free-space spectral filtering, and residual anomalous-dispersion fibre sections. Although built entirely from non-polarisation-maintaining fibres, the oscillator achieves stable fundamental-rate operation without requiring intentional nonlinear polarisation evolution or polarisation-sensitive mode-locking components.

In the stationary state, the laser operates not in a conventional coherent single-pulse mode-locked regime, but in a noise-like pulse regime. This is evidenced by the rectangular nanosecond-scale temporal envelope, the broad optical spectrum shaped by the offset FBGs, and the autocorrelation trace consisting of a broad pedestal with a narrow coherence spike. Real-time measurements further show that the cavity evolves from noise through transient multi-pulse states and pulse competition before stabilising as a single noise-like envelope at the fundamental repetition rate. Numerical simulations reproduce the same qualitative behaviour and likewise do not converge toward a coherent single-pulse solution.

The observed regime can be seen as a dissipative equilibrium between normal dispersion, SPM-induced spectral broadening, gain saturation, and spectral filtering. In this framework, nonlinear propagation widens the spectrum across the two offset FBG passbands, while the FBGs regularly truncate the broadened spectrum and introduce spectral loss on each round trip. Normal dispersion supports a heavily chirped broad envelope rather than soliton-like compression, and gain saturation prevents further growth. These processes collectively maintain a statistically steady envelope while allowing rapid fluctuations of sub-pulse structures within it, in line with the characteristic features of noise-like pulses (ref. 38-40).

Noise-like pulses are statistically stationary envelopes that contain rapidly fluctuating sub-pulses with short coherence times. Their formation cannot be attributed to a single universal mechanism (ref. 38, 39). In the case of Tm-doped fibre Mamyshev oscillators, an interplay of several effects leads to the stabilisation of a statistically stationary pulse envelope while permitting rapid fluctuations in internal structure. Mainly, the characteristic evolution of Mamyshev oscillators, in which SPM broadens the spectrum until it overlaps with the two offset FBG passbands while the filters truncate the broadened spectrum each round trip, supports a dissipative energy balance in strongly nonlinear normal-dispersion cavities. This intrinsic process stabilises a broad, energy-clamped envelope while still allowing stochastic sub-pulse structure (ref. 40).

Furthermore, rare-earth population dynamics can offer an additional means of nonlinear amplitude regulation in fibre lasers. Unlike standard Yb- and Er-doped fibre lasers, ion–ion energy-transfer processes are especially significant and, therefore, crucial in $Tm^{3+}$-doped fibres (ref. 35), since excited-state absorption, cross-relaxation, and upconversion-assisted dynamics can create distributed intensity-dependent loss. This loss contributes to peak-power clamping and can promote the formation of rapidly fluctuating sub-pulses within a broader envelope (ref. 39).

Stimulated Raman scattering can also serve as an energy-relief channel that promotes envelope stabilisation but decreases intrabunch coherence (ref. 39,41). Finally, in cavities containing non-PM fibres and polarisation-selective elements, nonlinear polarisation evolution may introduce an additional intensity-dependent transmission that could contribute to pulse bunching and noise-like pulse formation (ref. 42, 43). However, these effects are negligible in the presented case.

Beyond its role as a testbed for studying self-organisation at longer wavelengths, the demonstrated Tm-doped Mamyshev oscillator is attractive for applications that benefit from broadband, low-coherence emission in the SWIR. Relevant examples include speckle-free wide-field endoscopic illumination, low-coherence interferometry, optical reflectometry, and tomography, as well as for advanced photoacoustic and photoluminescence excitation experiments, e.g., three-photon absorption or plasma spectroscopy (ref. 38), data storage (ref. 44), and supercontinuum generation (ref. 45). More broadly, the minimal fibre-integrated architecture and deeper understanding of the underlying phenomena in Mamyshev oscillators can facilitate future extension towards mid-infrared operating wavelengths and integrated platforms (ref. 46).

In conclusion, we have achieved, for the first time to the best of our knowledge, self-starting pulse generation in the all-fibre, all-normal-dispersion Tm-doped Mamyshev oscillator at approximately 1.9 µm. By combining dispersion-engineered active and passive fibres with chirp-free broadband offset FBG filtering, the minimal three-element Fabry–Perot cavity evolves from noise to stable, noise-like pulse generation at the fundamental repetition rate without external seeding or active modulation. Real-time measurements and numerical modelling reveal a build-up pathway governed by stochastic background amplification, transient multi-pulsing, pulse competition, and the final stabilisation of a single noise-like envelope. These results offer an elegant platform for studying fundamental phenomena underlying ultrafast self-organisation and for developing practical low-coherence SWIR fibre sources.

## Methods

### Active fibre design and fabrication

The key laser component is a specially designed, normal-dispersion Tm-doped fibre based on a germanosilicate matrix. The fibre preform for the active fibre was fabricated using the modified chemical vapour deposition technique as described in (ref. 20). After depositing a buffer $SiO_2$–$P_2O_5$–F glass cladding with a refractive index similar to that of the silica support tube, several germanosilicate layers were deposited, with a gradual increase in $GeO_2$ concentration. Thulium was introduced using volatile organometallic thulium complex $Tm(tmhd)_3$ as a starting material. The $Tm^{3+}$ ion concentration of $6.2 \cdot 10^{25}$ m$^{-3}$ ensures approximately 21.5 dB m$^{-1}$ absorption at the pump wavelength. The maximum $GeO_2$ concentration in the core area reached 42 wt%. The active fibre has a 3.3 µm core diameter, an NA of 0.4, an effective mode field area of 14 µm$^2$, and a group velocity dispersion of 0.067 ps$^2$ m$^{-1}$ at the laser operating wavelength around 1.9 $\mu$m. The estimated effective Kerr coefficient of the fibre is 6 (W km)$^{-1}$.

### Near-zero normal-dispersion passive fibre design and fabrication

To enable efficient self-pulsation at the onset of pulse generation, the passive optical fibre must combine low normal group-velocity dispersion with high nonlinearity. This ensures sufficient spectral broadening via SPM across both filters, which is critical for initiating stable mode-locking. To achieve these properties, we designed and fabricated an optical fibre with a carefully tailored refractive index profile, resulting in near-zero normal dispersion.

The fibre core contains 38 mol% $GeO_2$, resulting in a 3.17% refractive index contrast between the core and cladding. The preform was drawn into fibre with a 3.2 µm core diameter and an ellipticity of approximately 3%, yielding a zero-dispersion wavelength of about 1.95 µm. The fibre shows low loss of 0.048 dB m$^{-1}$ at 1.98 µm, and a cut-off wavelength near 1.344 µm. Additionally, the developed fibre is compatible with commercially available normal-dispersion fibres with an effective mode field area of 18 µm$^2$ at the operational wavelength, enabling low splicing losses. The group velocity and third-order dispersion at 1.9 µm are only - 1 ps nm$^{-1}$ km$^{-1}$ and 0.056 ps$^3$ km$^{-1}$, respectively. The fibre is estimated to have an effective nonlinear Kerr coefficient of 6.0 (W km)$^{-1}$, based on an effective nonlinear refractive index $n_{2,\mathrm{eff}}$ of $3.0 \cdot 10^{-20}$ m$^2$W$^{-1}$.

### Filters design

A pair of top-hat FBGs with 4.2 and 3.7-nm bandwidth centred at 1.9 $\mu$m was inscribed in normal dispersion

UHNA1 fibre (from *Coherent*). The inscription was carried out using a 266 nm ultraviolet femtosecond laser in a Talbot interferometer with a phase mask having a pitch of 1379 nm. To achieve a broad bandwidth, four FBGs were superimposed one on top of another. The inscription process for each grating was continued until the set reflectivity value was reached. After that, the angular position of the interferometer mirrors was shifted for the inscription of the next grating (ref. 47). Such superimposed gratings benefit from a relatively broad bandwidth, short effective length, and negligible chirp, and therefore do not cause drastic variations to the pulse temporal profile.

**Experimental setup**

The schematic of the all-fibre Mamyshev Tm-doped fibre laser, shown in Fig. 1, consists of a 1.5-m long section of gain Tm-doped fibre positioned symmetrically between two ∼27-m long sections of nonlinear normal-dispersion fibre and a pair of superimposed FBGs. The gain fibre is pumped directly through one of the FBGs via an Er-doped master oscillator fibre amplifier (Box Optronics FLD-CH18-10SM-FA2 & Agiltron1C0112333), operating at 1.56 $\mu$m with maximum power of 4.1 W.

To control the mismatch of the filters, one of the FBGs was bonded to a flexible steel strip clamped between a fixed mount and a micrometre-driven translation stage. Advancing the stage bends the strip upwards or downwards, thereby stretching or compressing the FBG. Consequently, a strain-induced refractive-index variation via the elasto-optic effect and a direct change in the grating period through compression or stretching led to a ~26-nm blueshift or a 9-nm redshift, respectively, with ±50 pm accuracy and repeatability.

Although the laser cavity has a long net length of 112 m, the cavity dispersion is primarily determined by the properties of the active fibre, reaching 0.44 $ps^2$, while the net cavity nonlinearity attains a value of 0.66 rad $W^{-1}$.

To capture the desired buildup, the laser is initially set to produce a stable train of ultrashort pulses, then turn off. A 33-GHz, 100-GS/s digital storage oscilloscope (Tektronix DPO73304SX) and the 22-GHz photodetector (Discovery Semiconductors DSC2-30S) were employed to record the real-time spectral dynamics. The equipment was triggered to take a single measurement when the pump was reactivated and the signal reached the pulse peak (ref. 33). The evolution of the intensity temporal profile was recorded simultaneously using a 12-GHz photodetector (EOT ET5000f). The dispersive broadening to convert the temporal waveform of the generated pulses into the spectral profile was realised with a pair of chirped fibre Bragg gratings (CFBGs) connected consecutively via fibre circulators (ref. 48, 35), resulting in a total DFT line length of 8.4 m. The CFBGs (from Terraxion) were centred at 1905 nm and had a linear chirp of 24 nm over their length of 140 mm. Such a chirp ensures a group delay of approximately 100 ps/nm. Overall, the measurement setup provides a 15-ps temporal and a 0.1-nm spectral resolution (ref. 49). The averaged output spectra were recorded using an optical spectrum analyser (Yokogawa AQ6375B) and were employed for normalising the single-shot spectra. An autocorrelator with FROG function (Femtoeasy MS-ROC-2020) was utilised for pulse duration measurements.

**Numerical methods**

To model the electric field slowly varying envelope $A(z,t)$ evolution inside the laser, we have used the following generalised nonlinear Schrödinger equation, which has been integrated numerically using a standard split-step Fourier algorithm:

$$\frac{\partial A}{\partial z} = \frac{g(z)}{2}A + \frac{g(z)}{2\Omega_g^2}\frac{\partial^2 A}{\partial t^2} - i\frac{\beta_{2,g,p,NL}}{2}\frac{\partial^2 A}{\partial t^2} + i\gamma_{g,p,NL}|A|^2 A - \frac{\alpha_{p,NL}}{2}A \quad (1)$$

, where $z$ is the spatial coordinate along the cavity, $t$ is a temporal coordinate in a co-moving reference frame with the pulse. $\beta_2$ denotes group velocity dispersion, $\gamma$ -is the Kerr nonlinearity coefficient, $\alpha$ - is the loss coefficient. Subscripts *g*, *p*, and *NL* refer respectively to the gain fibre, the passive fibre with inscribed FBGs and the sections of nonlinear normal-dispersion fibre. The gain is defined by

$$g(z) = \frac{g_0}{1+(E_{sat})^{-1}\int_{t_\omega}|A(z,t)|^2 dt} \quad (2)$$

where $g_0$ denotes the unsaturated gain, $E_{sat}$ being the saturation energy, $\Omega_g$ is the half-width at half maximum bandwidth of the gain, and the integral is performed over the full simulation time window $t_\omega$.

The gain fibre was simulated using the following parameters: $\beta_{2,g}$ = 77 ps$^2$km$^{-1}$, $\gamma_g$ = 7.7 W$^{-1}$km$^{-1}$, $\Omega_g$ = $2\pi$·1.66rad/ps, $E_{sat}$ = 100 nJ, $g_0$ = 4.75 m$^{-1}$, $L_g$ = 1.5 m. The loss in active fibre can be neglected, i.e. $\alpha_g$ = 0

In the passive fibres, the gain is zero, i.e., $g$ = 0. Parameters values used in the simulation of pulse propagation in the sections of passive fibre are: $\beta_{2,p}$ = 40 ps$^2$km$^{-1}$, $\beta_{2,NL}$ = 1.7 ps$^2$km$^{-1}$, $\gamma_p$ = 2.4 W$^{-1}$km$^{-1}$, $\gamma_{NL}$ = 2.5 W$^{-1}$km$^{-1}$, $\alpha_{p,NL}$ = 0.075 dB/m. Following the laser system setup depicted in Fig. 1(a), the lengths of fibre sections with FBG and nonlinear fibre on the left side from the gain fibre are $L_{p1}$ = 0.3 m,$L_{NL1}$ = 28 m, respectively, and $L_{NL2}$ = 24 m, $L_{p2}$ = 0.3 m – to the right-hand side of the gain fibre (the output-side FBG).

The two cavity mirrors are defined by the following transfer function that acts multiplicatively on the Fourier transform of the field amplitude:

$$F_{1,2}(\omega) = r_{1,2}\exp\left(\frac{(\omega \pm \omega_f)^6}{2\sigma_{1,2}^6}\right) \quad (3)$$

with mirror reflectivities of $r_1$ = 0.95, $r_2$ = 0.85, spectral bandwidths of $\sigma_1$ = $2\pi \cdot 155$ rad/ns, $\sigma_2$ = $2\pi \cdot 136$ rad/ns, and frequency offset between the mirrors of $\omega_f$ = $2\pi \cdot 195$ rad/ns.

To phenomenologically simulate the fact that the final stage of the gain fibre behaves as a saturable absorber, given the fact that most the pump is absorbed to achieve population inversion in the first and main part, like in (ref. 50), the field amplitude has been multiplied by the following transfer function $\alpha = \alpha_{ns} + \alpha_0 / \left(1 + |A|^2 / P_{sat}\right)$ where $\alpha_0$= 0.15, $\alpha_{ns}$ = 0.85, and $P_{sat}$= 100 W.

## Acknowledgements

D.C.K.and M.C. acknowledge the support of the Deutsche Forschungs Gemeinschaft (DFG–German Research Foundation, Project No.CH26001-1). AMP, ARC, and MC acknowledge support from Engineering and Physical Sciences Research Council (Research Grant EP/Y001915/1). M.V.Y. acknowledges the support of the state assignment of the Russian Ministry of Science and Education (FFSR-2025-0005). The authors would like to acknowledge the group of Prof. Ole Bang from DTU Denmark for characterising the tailored fibres' dispersion.

## Author contributions

D.C. designed the laser cavity and performed the experiments. K.G. designed and fabricated the FBGs. A.R.C. assisted with laser generation stabilisation experiments. M.E.L. and S.S.A. made calculation of the gain fiber parameters. M.V.Y. developed the gain optical fibre. S.U., C.A., and A.L. designed, fabricated and characterised the nonlinear passive fibre. M.C. developed the research idea, contributed to the design, supervised the research, and analysed the results. All authors contributed to the preparation of the manuscript.

## Data availability

The data that support the findings of this study are openly available in Figshare at 10.6084/m9.figshare.32090344, reference number 32090344.

## Conflict of interest

The authors declare no competing interests.